\documentclass[reprint,amsmath,amssymb,aps,prl,superscriptaddress]{revtex4-2}
\usepackage{array}
\usepackage{amsmath}
\usepackage{bm}
\usepackage{color}
\usepackage{float}
\usepackage{graphicx}
\usepackage{hyperref}
\usepackage[normalem]{ulem}
\usepackage{comment}

\hypersetup{colorlinks=true,allcolors=blue}
\usepackage{natbib}
\usepackage{newtxtext}
\usepackage{newtxmath}
\usepackage[caption=false]{subfig}

\newcommand\J{\mathcal{J}}

\begin{document}

\title{Uncovering footprints of dipolar-octupolar quantum spin ice from neutron scattering signatures
}
\author{Masashi Hosoi}
\thanks{These authors contributed equally to this work.}
\affiliation{Department of Physics, University of Tokyo, 7-3-1 Hongo, Bunkyo, Tokyo 113-0033, Japan}
\author{Emily Z. Zhang}
\thanks{These authors contributed equally to this work.}
\affiliation{Department of Physics, University of Toronto, Toronto, Ontario M5S 1A7, Canada}
\author{Adarsh S. Patri}
\affiliation{Department of Physics, Massachusetts Institute of Technology, Cambridge, MA 02142, USA}
\author{Yong Baek Kim}
\affiliation{Department of Physics, University of Toronto, Toronto, Ontario M5S 1A7, Canada}


\begin{abstract}
Recent experiments on Ce$_2$Zr$_2$O$_7$ suggest that this material may host a novel form of quantum spin ice, a three-dimensional quantum spin liquid with an emergent photon. The Ce$^{3+}$ local moments on the pyrochlore lattice are described by pseudospin 1/2 degrees of freedom, whose components transform as dipolar and octupolar moments under symmetry operations. In principle, there exist four possible quantum spin ice regimes, depending on whether the Ising component is in the dipolar/octupolar channel, and two possible flux configurations of the emergent gauge field. In this work, using exact diagonalization and molecular dynamics, we investigate the equal-time and dynamical spin structure factors in all four quantum spin ice regimes using quantum and classical calculations. Contrasting the distinct signatures of quantum and classical results for the four possible quantum spin ice regimes and elucidating the role of quantum fluctuations, we show that the quantum structure factor computed for the $\pi$-flux octupolar quantum spin ice regime is most compatible with the neutron scattering results on Ce$_2$Zr$_2$O$_7$.
\end{abstract}

\maketitle

\textit{Introduction.}--- Experimental identification of quantum spin liquids is an outstanding issue in quantum condensed matter physics. In principle, detection of fractionalized quasiparticles and emergent gauge fields would be a direct confirmation of quantum spin liquids\cite{Balents2010c,Knolle2019,Savary2017c,Zhou2017b,Broholm2020}. However, such signatures in conventional spectroscopic probes may not have sufficiently sharp features, and 
competing quantum spin liquid phases may also exist\cite{Dodds2013,Desrochers2021}. 
This difficulty may be circumvented if one can experimentally determine the parameters of the underlying microscopic model, which may be used to make clear theoretical predictions for competing quantum spin liquids.

Recent experiments\cite{Gaudet2019,Gao2019,Sibille2020,Smith2021} and theoretical analyses\cite{Yao2020a, Bhardwaj2021} on Ce$_2$Zr$_2$O$_7$ have made important progress in this regard. 
In Ce$_2$Zr$_2$O$_7$, the local moments on Ce$^{3+}$ ions form the pyrochlore lattice\cite{Sibille2015}. The lowest Kramers doublet of the Ce$^{3+}$ ion can be described by the pseudospin-1/2 degrees of freedom, {\bf S}, which transform as dipolar ($S^x, S^z$) and octupolar ($S^y$) moments defined in terms of the local quantization axis at each site\cite{Rau2019,Huang2014}. 
In the local frame, the effective Hamiltonian is an $XYZ$ model written in terms of three coupling constants, ${\cal J}_x, {\cal J}_y, {\cal J}_z$\cite{Rau2019,Huang2014}, which
describe the interactions between $S^{\alpha}$ ($\alpha=x,y,z)$ components at nearest-neighbor sites.
Such systems can host quantum spin ice (QSI) states, especially near the Ising limit, where one of the interactions is antiferromagnetic and dominates over the other two\cite{Hermele2004a,Ross2011,Savary2012,Lee2012,Gingras2014,Li2017a,Savary2021}. 

When the dominant Ising interaction is in the octupolar (dipolar) channel or $S^y$ ($S^z$ or $S^x$) component, the corresponding QSI state is called the octupolar (dipolar) QSI, dubbed O-QSI (D-QSI). Near the Ising limit, the low energy theory can be described by an emergent lattice $U(1)$ gauge theory,
\begin{equation}\label{heff}
H_{\rm eff} = U {\bf E}^2 + K \cos (\nabla \times {\bf A}), 
\end{equation}
where {\bf E} and {\bf A} are the emergent electric and gauge fields, and are related to the Ising and transverse components of the pseudospin-1/2, respectively\cite{Lee2012,Chen2017a,Benton2018}.
Here, $K < 0$ ($K > 0$) for ferromagnetic (antiferromagnetic) transverse interactions and $U$ is taken to be a large constant. These regimes are often called unfrustrated ($K < 0$) and frustrated ($K > 0$) regimes, respectively. 
When $K < 0$, the existence of the deconfined phase with fractionalized spinons and propagating emergent photons is rigorously established by quantum Monte Carlo simulations\cite{Shannon2012a, Benton2012a}. 
Meanwhile, the situation is less certain in the frustrated regime, where quantum Monte Carlo (QMC) suffers from the famous sign problem. In this regime, only exact diagonalization (ED) studies on small system sizes are available and suggest the existence of a quantum spin liquid\cite{Patri2020,Benton2020}, but the nature of the state is not well understood. 
The argument $\nabla \times {\bf A}$ in \eqref{heff} represents the emergent gauge flux through a hexagonal loop in the underlying pyrochlore lattice, and $K < 0$ ($K > 0$) would favor zero ($\pi$) flux configurations in the corresponding ground states\cite{Hermele2004a, Lee2012, Savary2012}. Therefore, there are four possible QSI states, namely zero-flux O-QSI (0-O-QSI) and zero-flux D-QSI (0-D-QSI) in the unfrustrated regime, and $\pi$-flux O-QSI ($\pi$-O-QSI) and $\pi$-flux D-QSI ($\pi$-D-QSI) in the frustrated regime. 

Recent experiments comparing data with classical molecular dynamics (MD) computations determined the system to be in the $\pi$-O-QSI regime, i.e. ${\cal J}_y > 0$ (the interaction between octupolar $S^y$) is the dominant Ising coupling, and the transverse couplings (${\cal J}_x, {\cal J}_y > 0$) are antiferromagnetic\cite{Gaudet2019,Gao2019,Sibille2020,Smith2021}. 

Here, we critically examine this proposal by theoretically investigating neutron scattering signatures of four possible QSI regimes. We use both ED studies on the quantum model\cite{KAWAMURA2017180} and MD on the classical model\cite{Conlon2009,Samarakoon2017,Zhang2019} to unveil distinct signatures in the equal-time static and inelastic dynamical structure factors for four different QSI regimes, and we compare the results with the experimental findings. We find that the signatures at $[hhl]=[001]$ and $[003]$ in the equal-time static structure factor and the high intensity at the $X$ point in the inelastic dynamical structure factor obtained in experiments can be identified in the quantum calculations in the $\pi$-O-QSI regime using only nearest-neighbour interactions. These features are absent in other three QSI states. 
Moreover, most of these features are not clearly seen in the classical Monte Carlo and MD simulations for the corresponding classical model, even in the $\pi$-O-QSI regime. Hence, these are quantum effects that are unique to the $\pi$-O-QSI regime. 

\begin{figure}[t!]
\includegraphics[width=0.7\columnwidth]{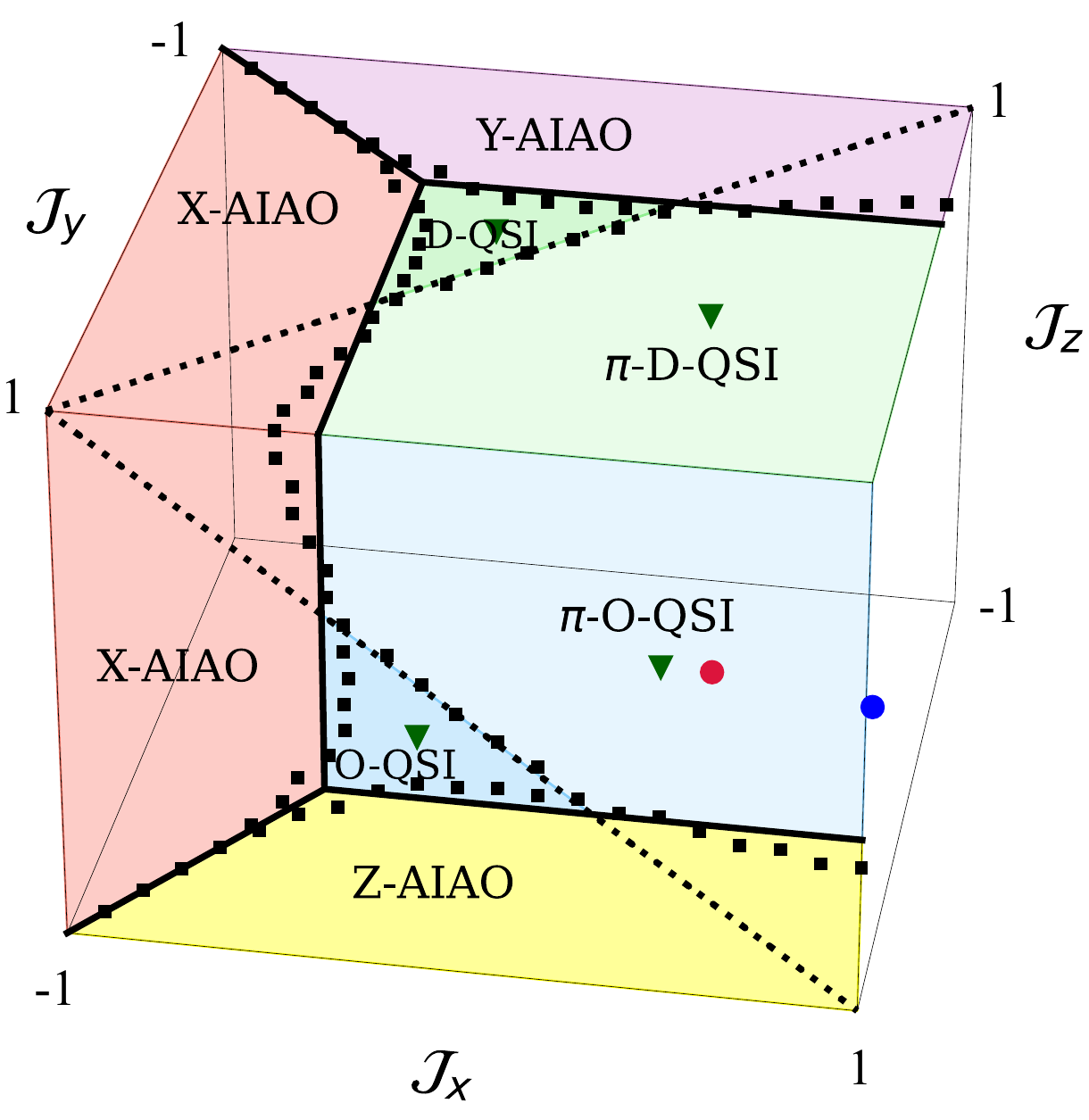}
\caption{\label{pdfig} Phase diagram of DO systems in zero magnetic field in the octupolar dominant and dipolar dominant regimes (See Supplemental Material \cite{SM} for the cube faces placed adjacent to each other). 
The depicted phases are on the Ising planes of $\mathcal{J}_{x,y,z} = 1$ respectively.
The X,Y,Z all-in, all-out phases are labelled as X,Y,Z-AIAO, while the $0-$flux ($\pi-$flux) quantum spin ice phases for the corresponding octupolar and dipolar dominant regimes are labelled as O-QSI ($\pi-$O-QSI) and D-QSI ($\pi-$D-QSI). The solid lines indicate the classical phase diagram phase boundaries, while the black squares are the 16-site ED phase boundaries from Ref. \citep{Patri2020}. The blue and red circles are superimposed experimentally extracted parameter sets from Ref. \citep{Smith2021} and \citep{Bhardwaj2021}, respectively.
The inverted-green triangles are the parameter sets employed in the MD and ED studies in the subsequent sections.
The dashed lines indicate the crossover from unfrustrated to frustrated exchange couplings.}   
\end{figure}

\textit{Microscopic Model of Dipolar-Octupolar Compounds.}--- 
In the pyrochlore compound, Ce$_2$Zr$_2$O$_7$, 
the lowest-lying doublet of ground states 
of the Ce$^{3+}$ ion
support non-trivial dipolar-octupolar (DO) moments\cite{Huang2014,Sibille2015,Gaudet2019,Gao2019}, which can be 
efficiently captured by pseudospin-1/2 operators, $\bf{S}$, where $S^{x,z}$ transform as dipolar moments $J_z$, while $S^{y}$ component transforms as the octupolar moment $J_y^3 - \overline{J_x J_x J_x}$, where the overline indicates a symmetrized product.

From the underlying symmetry of the pyrochlore lattice, the nearest-neighbour pseudospin Hamiltonian for DO compounds subjected to a perturbatively weak applied magnetic field, $\bm{h}$, is of the form,
\begin{align}
\label{eq_XYZh}
  H_{\mathrm{XYZh}} &=\sum_{\langle i,j\rangle} \mathcal{J}_\mu {S}_i^\mu {S}_j^\mu -\sum_i \Bigg[({\bm h}\cdot\hat{z}_i)  \left({g}_x {S}_i^x+ {g}_z {S}_i^z \right) \Bigg] ,
\end{align}
where the repeated indices sum over  $\{x,y,z\}$, $\hat{z}_i$ is the local-$z$ direction of the sublattice at site $i$.
We note that this model is written in a rotated-local basis. Since both the $S^{x,z}$ transform as microscopic dipolar moments, they are permitted to couple by the linear anisotropic Zeeman term, while the octupolar $S^y$ moment requires a cubic-in-$\mathbf{h}$ and is neglected in the perturbatively weak magnetic field limit.
Moreover, though both $S^{x,z}$ transform as dipolar moments, we note that in Ce$_2$Zr$_2$O$_7$, $S^x$ is ultimately a microscopic octupolar moment ($J_x^3-\overline{J_xJ_yJ_y}$ that belongs to the same irrep as the dipolar moment $J_z$), and as such we assume that the $g_x$ factor to zero for simplicity; indeed, this assumption amounts to taking negligible mixing between the $S^{x,z}$ moments.

\textit{Phase Diagram and Microscopic Parameters.}--- 
The pseudospin exchange model of Eq. \ref{eq_XYZh} in the absence of an external magnetic field can be solved either by the Monte Carlo simulations for the classical model and ED on finite-size clusters for the quantum model\cite{Patri2020,Benton2020}.
Depending on the dominant pseudospin exchange constant $\mathcal{J}_{\mu}$, the phases are classified as dipolar-dominant ($\mathcal{J}_z=1$ or $\mathcal{J}_x=1$) or octupolar-dominant ($\mathcal{J}_y=1$) regimes. The resulting phase diagrams are shown in Fig. \ref{pdfig}.
The classical phase boundaries are denoted by the solid lines, while the 16-site ED phase diagram lines are indicated by the solid squares.
In addition to the symmetry-broken all-in, all-out phases (AIAO, where all the pseudospins on a given sublattice are pointing into, or out of a tetrahedron), there exist two quantum-disordered paramagnetic phases, where two different QSI states labelled as $0$-QSI and $\pi$-QSI (with the appropriate dipolar/octupolar name) are supposed to occur.
We note that though the classical calculation is not receptive to the differences between the frustrated ($\pi$-flux) and un-frustrated ($0$-flux) QSI regimes, the ED calculations on the quantum model show a distinguishing phase boundary. 
Superimposed on the theoretical phase diagram in the octupolar quadrant are the parameter choices (red and blue dots) extracted from the recent experimental studies\cite{Bhardwaj2021,Smith2021}.
In order to investigate
the distinctive physical responses of the four different QSI regimes, 
we choose a representative parameter set (depicted by green-inverted triangles in the phase diagram) for each QSI regime and compute the classical and quantum structure factors. 

\begin{figure*}[t!]
\includegraphics[width=2.1\columnwidth]{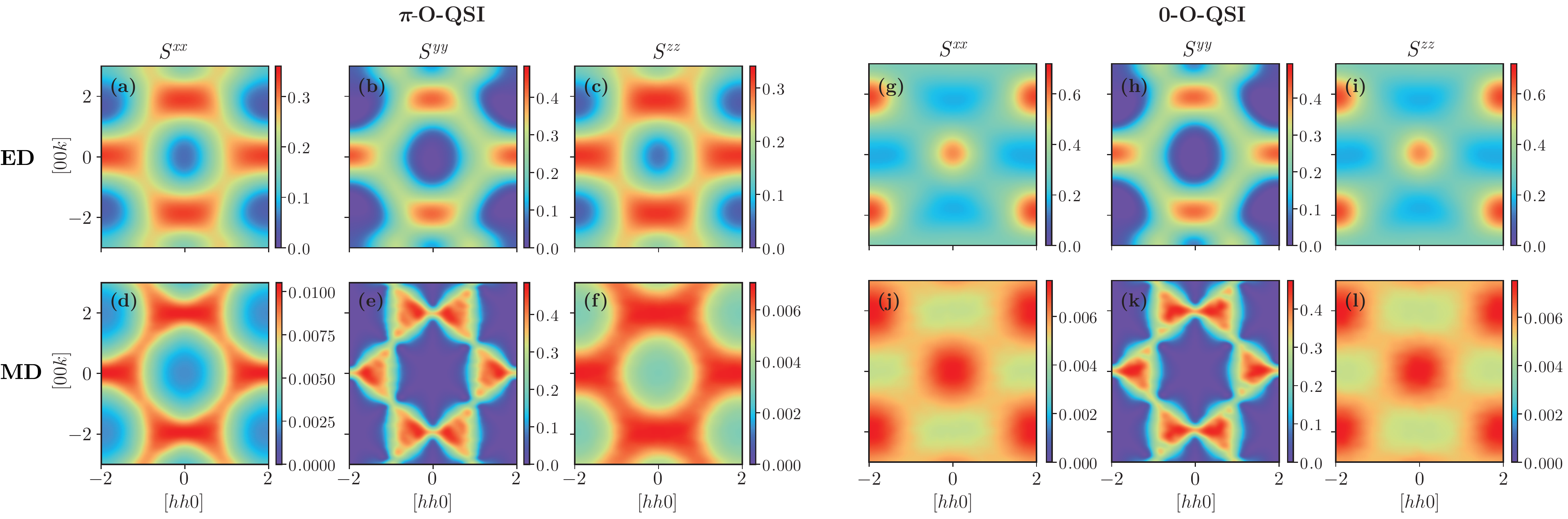}
\centering
\caption{\label{octopolar} Diagonal pseudospin correlations for the octopolar quantum spin ice (O-QSI) states in the $[hhk]$ plane. (a)-(c) and (g)-(f) were obtained from 32-site ED, and (d)-(f) and (j)-(l) were obtained using classical MD simulations. The $\pi$-O-QSI correlations (a)-(f) were computed using the parameter set $(\J_x, \J_y, \J_z)=(0.5,1.0,0.25)$, whereas the $0$-O-QSI correlations (g)-(l) used $(\J_x, \J_y, \J_z)=(-0.1,1.0,-0.1)$.}
\end{figure*}
\begin{figure*}[t!]
\includegraphics[width=2.1\columnwidth]{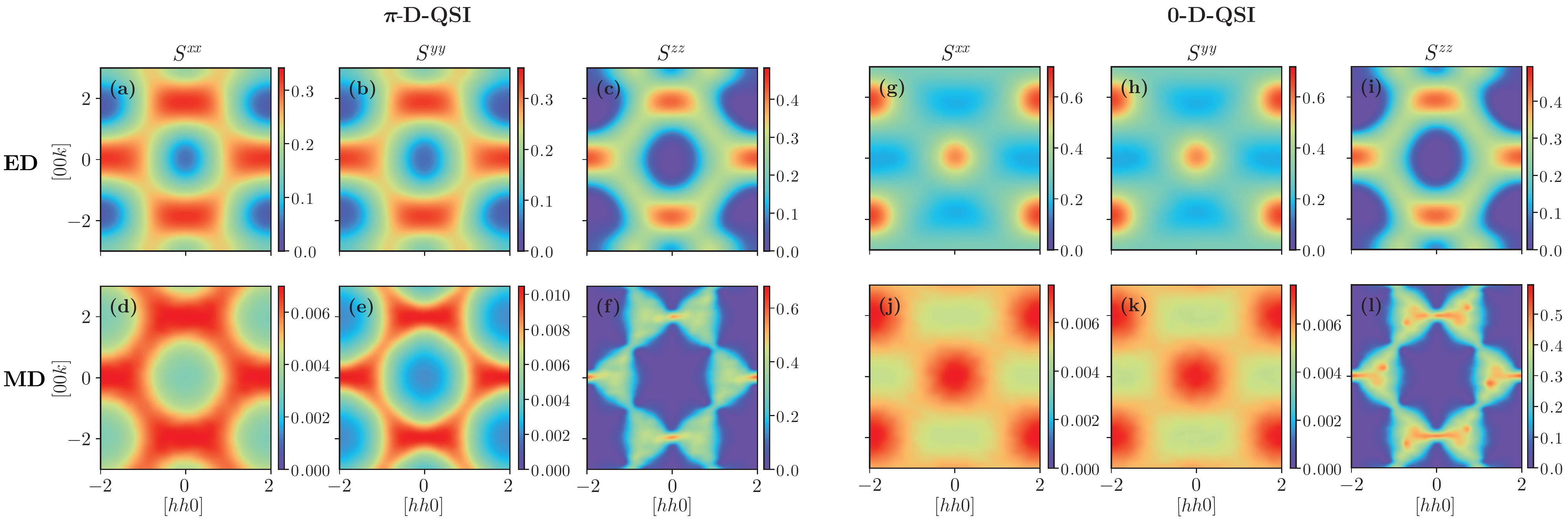}
\centering
\caption{\label{dipolar} Diagonal pseudospin correlations for the dipolar quantum spin ice (D-QSI) states in the $[hhk]$ plane. (a)-(c) and (g)-(f) were obtained from 32-site ED, and (d)-(f) and (j)-(l) were obtained using MD simulations. The $\pi$-D-QSI correlations (a)-(f) were computed using the parameter set $(\J_x, \J_y, \J_z)=(0.25,0.5,1.0)$, whereas the $0$-D-QSI correlations (g)-(l) used $(\J_x, \J_y, \J_z)=(-0.1,-0.1,1.0)$.}
\end{figure*}

\textit{Equal-time pseudospin correlations.}--- 
The neutron scattering intensity profiles provide remarkably discriminating features for the various types of QSIs due to the multipolar nature of the underlying moments.
In particular, the purely octupolar $S^{y}$ moment fails to generate appreciable contributions to the scattering intensity (due to the dominant dipolar coupling to the neutron's magnetic moment), and neither do the $S^x$-involved correlation functions (due to the underlying assumption of a negligible $x$-component of the $g$-tensor).
As such, the primary instigator in the scattering intensity is the correlation function associated with $S^z$ moment, indicating that states possessing the same $S^{zz}$ correlation have the same 
scattering response. 

\begin{figure*}[t!]
\centering
\includegraphics[width=2\columnwidth]{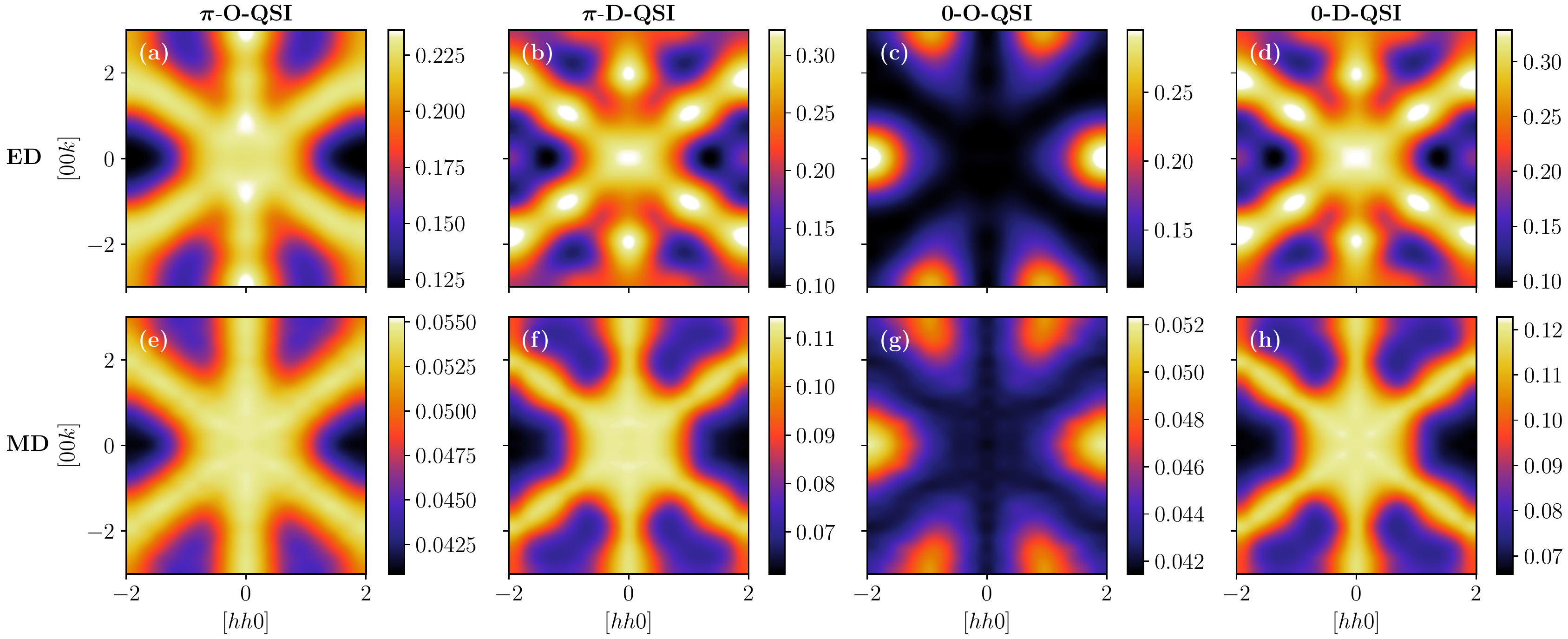}
\caption{\label{NS} Neutron scattering equal-time structure factor $S(\mathbf{q})$ for the octopolar and dipolar spin liquid regimes. (a)-(d) were obtained used 32-site ED, whereas (e)-(h) were computed using MD simulations. The contrast between the uniform rod-like signatures in the classical calculations (e) are contrasted with the high-intensity peaks in its quantum counterpart (a) for the $\pi$-O-QSI regime.  }
\end{figure*}

We present the equal-time diagonal pseudospin correlations, defined as $S^{\mu\mu}({\bm q})=\frac{1}{N}\sum_{i,j}e^{-\mathrm{i}{\bm q}\cdot({\bm R}_i-{\bm R}_j)}\langle S_i^\mu S_j^\mu\rangle$, in the $[hhk]$ momentum plane for the O-QSI regimes in Fig. \ref{octopolar}, and the D-QSI regimes in Fig.  \ref{dipolar}. 
The classical and quantum phases exhibit some marked similarities and important differences.
Since the 32-site ED computations can only access a finite number of momentum positions, we use an interpolation scheme to plot the quantum results for better comparison with the classical computations. Details of the interpolation scheme used can be found in \citep{SM}.
Firstly, the $S^{xx}$ and $S^{zz}$ correlations in Fig. \ref{octopolar}, and $S^{xx}$ and $S^{yy}$ in Fig. \ref{dipolar} possess the same qualitative features in both classical and quantum cases.
In contrast, the correlation functions associated with the Ising sectors, $S^{yy}$ ($S^{zz}$) in Fig.\ref{octopolar} (\ref{dipolar}), exhibit sharp differences.
The correlations display the characteristic pinch-point features of classical spin-ice-like phases, while these become washed out in their quantum counterparts. This effect is similar to the difference in the equal-time structure factors between the classical spin ice and quantum spin ice, computed in the unfrustrated regime by classical and QMC, respectively\cite{Benton2012a}.
Therefore, this difference mainly comes from quantum fluctuations even though some finite size effects are also present.

Equipped with the correlation functions, we present the equal-time neutron scattering structure factor,
\begin{equation}
  S({\bm q})=\frac{1}{N}\sum_{i,j}\left[\hat{z}_i\cdot\hat{z}_j-\frac{(\hat{z}_i\cdot{\bm q})(\hat{z}_j\cdot{\bm q})}{q^2}\right]e^{-\mathrm{i}{\bm q}\cdot({\bm R}_i-{\bm R}_j)}\langle S_i^z S_j^z\rangle,
\end{equation}
for all four spin ice regimes in Fig. \ref{NS}. Here, $\hat{z}_i$ is the local $z$-axis at site $i$.
As discussed above, due to the multipolar nature of the microscopic moments, the dominant contribution to the scattering intensity arises from the $S^{zz}$ correlation functions.
Remarkably, in the octupolar-dominant phases, the $S^{zz}$ correlation functions for the 0-flux and $\pi$-flux QSI phases are markedly distinct.
This noticeable difference carries over to their respective (both quantum and classical) neutron scattering intensities in Fig. \ref{NS}(a,e) and \ref{NS}(c,g).
Both phases possess rod-like motifs, but their scattering intensities are reversed.
More importantly, the $\pi$-O-QSI uniquely possesses high-intensity peaks at the $[001]$ and $[003]$ locations in the quantum calculation;
this feature is absent in the other QSI phases (as well as in the classical simulations).
The enhanced intensity at $[001]$ and $[003]$, as well as the rod-like motif, provide clear demarcating signatures for identifying the $\pi$-flux O-QSI phase. The uniqueness of the scattering signatures for the $\pi$-O-QSI phase persists even in the spin-flip (SF) and non-spin-flip (NSF) channels. In the ED calculations for $\pi$-O-QSI in the NSF channel, there is a modulated intensity structure similar to the experimental data \cite{Smith2021}, even though our model only considers nearest-neighbour interactions. In contrast, the ED calculation for 0-O-QSI in the NSF channel shows no such correlations, and the corresponding classical models show no modulation patterns either. Hence, by comparing the contrasting results from the quantum models, we conclude that this modulation in the $\pi$-O-QSI phase in the NSF channel arises from quantum fluctuation effects. More details on the spin-polarization analysis can be found in the Supplemental Material\cite{SM}. 
For the D-QSI phases, the $S^{zz}$ correlation functions are qualitatively similar for both the 0-flux and $\pi$-flux phases, which expectedly carries over to the scattering intensity in Fig. \ref{NS}(b,f) and (d,h).
Though the D-QSI and O-QSI possess 
rod-like features, the explicit pinch-point features of the classical D-QSI phase (smeared out in the quantum cases) provides a further
distinguishing feature between the two multipolar families.

\begin{figure*}[t!]
\includegraphics[width=2\columnwidth]{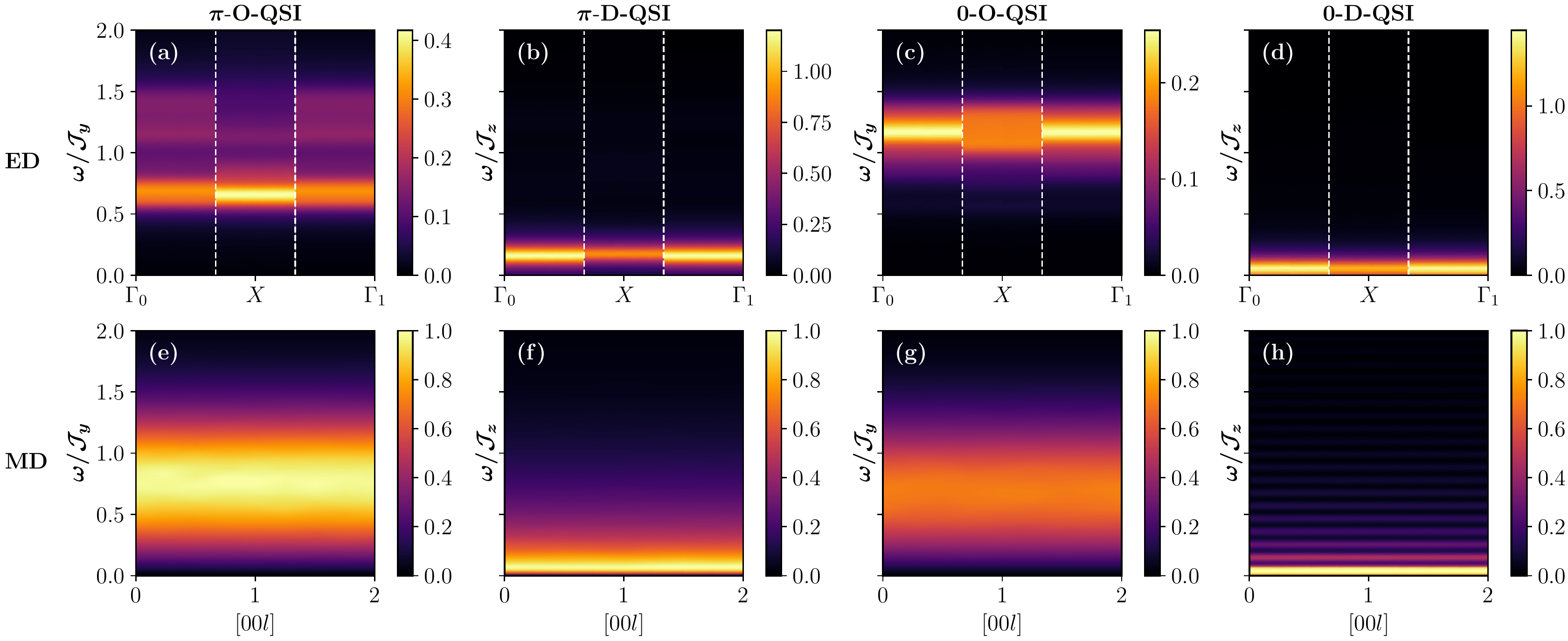}
\caption{\label{dynamical} Dynamical spin structure factor as a function of energy and momentum along the $[00l]$ momentum cut for each quantum spin liquid regime. (a)-(d) depict the results obtained from 32-site ED, and (e)-(h) the results from MD. The classical results were multiplied by a factor of $\beta\omega$, where $T=0.06$ in units of $|\mathcal{J}_y|$ or $|\mathcal{J}_z|$. The intensities of the (e)-(h) are given in arbitrary units, with (g) normalized to the max. amplitude of (e), and (h) normalized to (f).  }
\end{figure*}

\textit{Dynamical Structure Factor.}--- 
Next, we turn to the energy dependence of the dynamical spin structure factor (DSSF) along the $[00l]$ momentum path for each of the four QSI regimes, shown in Fig. \ref{dynamical}. We first observe 
a very weak momentum dependence in the classical data (Figs. \ref{dynamical}(e)-(h)), whereas the quantum results show clear nonuniform intensity distribution along this path. Notice that the 32-site ED computations can only access the following momentum positions along this path, namely $\Gamma_0$: $[hhk]=[000]$, $X$: $[hhk]=[001]$, and $\Gamma_1$: $[hhk]=[002]$.

We draw attention to a crucial feature of the quantum data, i.e. the $\pi$-O-QSI regime exhibits high-intensity at the $X$ point, seen in Fig. \ref{dynamical}(a). This may be consistent with the high intensity peak at $[001]$ in the equal-time neutron structure factor provided in the previous section. In contrast, the other three cases shown in Figs. \ref{dynamical}(b)-(d) show low-intensity troughs at this point. This peak in the DSSF at the $X$ point is therefore unique to the $\pi$-O-QSI in comparison to the other three states.

As shown in Fig. \ref{dynamical}(a), the DSSF in the $\pi$-O-QSI regime displays a gapped excitation spectrum with a gap of $\Delta/\mathcal{J}_y\sim0.6$. 
If we interpret this result assuming that the ground state is indeed the $\pi$-O-QSI, then the gapped excitations would correspond to spinon excitations with a quantum-renormalized excitation gap. Recall that the dominant contribution to the DSSF comes from the dipolar sector $S^z$ and the octupolar Ising sector, carrying the information about the emergent photon, does not contribute in the leading order neutron scattering amplitude as explained earlier.

\textit{Discussion.}--- 
In this paper, we investigated neutron scattering signatures of quantum and classical models of four different QSI regimes. Based on both the equal-time and the dynamical spin structure factors, we conclude that the quantum model in the $\pi$-O-QSI regime is the most consistent with experimental observations\cite{Smith2021}. Specifically, the high-intensity peaks at the $[001]$ and $[003]$ points measured in elastic neutron scattering, and the high-intensity peak at the $X$ point in inelastic neutron scattering measurements are both present in our results for $\pi$-O-QSI obtained from 32-site ED. It is also intriguing that in contrast to previous theoretical studies \cite{Bhardwaj2021}, where these features only clearly became apparent in the classical treatment when including next-nearest-neighbour couplings, the aforementioned discriminating features occur within the framework of nearest neighbour exchange interactions in our ED studies. Indeed, due to the localized nature of the $f$-electron orbitals, one naturally expects that the dominant exchange couplings are first nearest-neighbour. This expectation is consistent with our quantum calculations, and highlights the importance of quantum fluctuations in this family of systems. 
We also note that the $0$-O-QSI regime can be excluded from consideration, since the intensity profile is opposite to what has been observed in experiments. 

Encouragingly, our quantum simulations imply that the model parameters obtained from previous experimental and classical model fits are able to produce the features measured in neutron scattering experiments. These parameters are within the regime where the $\pi$-O-QSI state is expected to occur, based on the phase diagram mapped out from classical and ED studies\cite{Patri2020, Benton2020}. Whether the $\pi$-O-QSI is the true ground state of Ce$_2$Zr$_2$O$_7$, however, requires further theoretical investigation. The equal-time and dynamical spin structure factors have not yet been analytically computed for the $\pi$-O-QSI state, hence the direct comparison with the numerical computations is not possible. Furthermore, there may exist other competing quantum spin liquid phases that show similar scattering features in the same parameter regime of the phase diagram\cite{Desrochers2021}. These issues will have to be settled in order to conclude that the $\pi$-O-QSI is indeed observed in the experiments.\nocite{Frommer2003,Moessner1998,Lakshmanan2011,Momma2011,Isakov2005}

\begin{acknowledgements}
We thank F\'elix Desrochers, Bruce Gaulin, and Hitesh Changlani for helpful discussions. We acknowledge support from the Natural Sciences and Engineering Research Council of Canada (NSERC) and the Centre of Quantum Materials at the University of Toronto. The 32-site ED calculations were done using a quantum lattice model solver H$\Phi$\cite{KAWAMURA2017180}. The computations for ED and MD were performed on the Cedar cluster, which is hosted by WestGrid and SciNet in partnership with Compute Canada. M.H. was supported by the Japan Society for the Promotion of Science through the Research Fellowship for Young Scientists and the Program for Leading Graduate Schools (MERIT) and grant number JP20J10725.
\end{acknowledgements}
\bibliography{U1-Pi}
\end{document}


\title{Supplementary Material}

\author{Masashi Hosoi}
\thanks{These authors contributed equally to this work.}
\affiliation{Department of Physics, University of Tokyo, 7-3-1 Hongo, Bunkyo, Tokyo 113-0033, Japan}
\author{Emily Z. Zhang}
\thanks{These authors contributed equally to this work.}
\affiliation{Department of Physics, University of Toronto, Toronto, Ontario M5S 1A7, Canada}
\author{Adarsh S. Patri}
\affiliation{Department of Physics, Massachusetts Institute of Technology, Cambridge, MA 02142, USA}
\author{Yong Baek Kim}
\affiliation{Department of Physics, University of Toronto, Toronto, Ontario M5S 1A7, Canada}

\maketitle

\setcounter{equation}{0}
\setcounter{figure}{0}
\setcounter{table}{0}

\renewcommand{\thesection}{S\arabic{section}}
\renewcommand{\theequation}{S\arabic{equation}}
\renewcommand{\thefigure}{S\arabic{figure}}
\renewcommand{\thetable}{S\arabic{table}}

\onecolumngrid

\section{Details of the Numerical Methods}\label{numerical}
\subsection{Exact Diagonalization}\label{ED}
There are only a few useful unbiased methods of quantum analysis for the frustrated parameter regions. For example, the quantum Monte Carlo simulation is not applicable in the frustrated regime due to the notorious negative-sign problem. Here, we employ the exact diagonalization (ED) method on a 32-site cluster. Note that we restrict our discussion to zero temperature.

To begin, we examine the diagonal local pseudospin correlations to capture the spin-ice signatures of the four different QSI regimes,
\begin{equation}
  S^{\mu\mu}({\bm q})=\frac{1}{N}\sum_{i,j}e^{-\mathrm{i}{\bm q}\cdot({\bm R}_i-{\bm R}_j)}\langle S_i^\mu S_j^\mu\rangle,
\end{equation}
where $N=32$ is the total number of sites, $i$ and $j$ are the site indices, ${\bm R}_i$ is the site location, and $\mu=x,y,z$ denotes the local basis.  As mentioned above, the $S^y$ component is considered the octupolar moment, thus it does not linearly couple to the neutron moment. The dipolar components are therefore the dominant contributions to the neutron scattering, and we ignore any higher order couplings to the octupolar moment. As observed in recent experiments \cite{Gao2019,Gaudet2019}, the $x$ component of the $g$-tensor is negligible as compared to the $z$ component (since mixing between $x$ and $z$ components of pseudospins is assumed to be almost zero). Therefore, the observable equal-time static neutron scattering structure factor is given by,
\begin{equation}
  S({\bm q})=\frac{1}{N}\sum_{i,j}\left[\hat{z}_i\cdot\hat{z}_j-\frac{(\hat{z}_i\cdot{\bm q})(\hat{z}_j\cdot{\bm q})}{q^2}\right]e^{-\mathrm{i}{\bm q}\cdot({\bm R}_i-{\bm R}_j)}\langle S_i^z S_j^z\rangle.
\end{equation}
Here, $\hat{z}_i$ is the local $z$-axis at site $i$. 
In order to capture the dynamical properties of the system, we also calculate the following dynamical structure factor,
\begin{equation}
\begin{aligned}
    S({\bm q},\omega)&=\frac{1}{2\pi N}\sum_{i,j}\int_{-\infty}^\infty dt\left[\hat{z}_i\cdot\hat{z}_j-\frac{(\hat{z}_i\cdot{\bm q})(\hat{z}_j\cdot{\bm q})}{q^2}\right]e^{-\mathrm{i}{\bm q}\cdot({\bm R}_i-{\bm R}_j)+\mathrm{i}\omega t}\langle S_i^z(t) S_j^z(0)\rangle\\
    &=\frac{1}{2\pi}\int_{-\infty}^\infty e^{\mathrm{i}\omega t}(\langle \mathcal{A}_{-{\bm q}}^\mu(t)\mathcal{A}_{{\bm q}}^\mu(0)\rangle-
    \langle \mathcal{B}_{-{\bm q}}(t)\mathcal{B}_{{\bm q}}(0)\rangle),
\end{aligned}
\end{equation}
where we employ the Einstein summation notation for $\mu=x,y,z$ and define $\mathcal{A}_{\bm q}^\mu=N^{-1/2}\sum_i e^{\mathrm{i}{\bm q}\cdot {\bm R}_i}\hat{z}_i^\mu S_i^z$ and $\mathcal{B}_{\bm q}=N^{-1/2}\sum_i e^{\mathrm{i}{\bm q}\cdot {\bm R}_i}(\hat{z}_i\cdot{\bm q}/q) S_i^z$. Using the fluctuation-dissipation theorem which relates the correlation functions and the retarded Green's function, we obtain,
\begin{equation}
    S({\bm q},\omega)=-\frac{1}{\pi}\mathrm{Im}\left[\langle \Phi_0|\mathcal{A}_{\bm q}^{\mu\dagger}
    \frac{1}{\omega+\mathrm{i}\delta-H}
    \mathcal{A}_{\bm q}^\mu|\Phi_0\rangle-\langle \Phi_0|
    \mathcal{B}_{\bm q}^{\mu\dagger}
    \frac{1}{\omega+\mathrm{i}\delta-H}
    \mathcal{B}_{\bm q}^\mu|\Phi_0\rangle\right],
\end{equation}
where $|\Phi_0\rangle$ denotes the ED ground state, and each term can be calculated by employing the shifted Krylov method \cite{Frommer2003}. 

\subsection{Molecular Dynamics}\label{MD}
To examine the classical limit, we employ finite temperature Monte Carlo (MC) techniques to obtain the spin correlations\cite{Moessner1998,Zhang2019}. Firstly, we use a combination of simulated annealing and parallel tempering for $5\times 10^6$ MC sweeps on a cluster of $4\times 8\times 8 \times 8$ sites. Fixing the magnitude of the classical spins $\mathbf{S}=(S_x, S_y, S_z)$ to be $S=1/2$, we allow the system to thermalize to a temperature of $T=0.06\ |{\cal J}|$, where ${\cal J}={\cal J}_y$ in the octopolar case, and ${\cal J}={\cal J}_z$ for the dipolar case. Then, we perform another $10^6$ MC sweeps, with measurements recorded every 10 steps.

We use the method of classical Molecular Dynamics (MD) to capture the dynamics of these systems\cite{Moessner1998,Zhang2019}. The spins are evolved according to the semi-classical Landau-Lifshitz-Gilbert equations of motion \cite{Lakshmanan2011},
\begin{align}
    \frac{d}{dt}\mathbf{S}_i = - \mathbf{S}_i\times \frac{\partial H}{\partial\mathbf{S}_i},
\end{align}
and we allow the system to evolve for a long but finite time of $t=60/|{\cal J}|$, with step sizes of $\delta t=0.05/|{\cal J}|$ to obtain $S_i^\mu (t) S_j^\nu (0)$. This process is repeated for every measurement from the MC simulations, and the results are averaged over to yield $\langle S_i^\mu (t) S_j^\nu (0)\rangle$. Finally, we numerically integrate over these both the classical and quantum time-evolved spins to obtain the energy- and momentum-dependent dynamical structure factors, $\mathcal{S}(\mathbf{q},\omega)$.
We recall that the dynamical structure factors in the quantum and classical cases have their own fluctuation-dissipation theorems i.e.
$\mathcal{S}_{\text{quantum}}(\mathbf{q},\omega) (1 - e^{-\beta \omega} ) = \chi''(\mathbf{q},\omega)$, and $\mathcal{S}_{\text{classical}}(\mathbf{q},\omega) = \beta \omega \chi''(\mathbf{q},\omega)$, where $\beta = 1/k_B T$ and $\chi(\mathbf{q}, \omega)$ is the dynamical susceptibility.
Under the framework of linear spin-wave theory, the dynamical susceptibilities appearing in both the quantum and classical cases are equivalent \cite{Zhang2019}, thus relating a classical dynamical structure factor to its quantum counterpart via a simple rescaling factor (in the limit of $T \rightarrow 0$) $\mathcal{S}_{\text{classical}}(\mathbf{q},\omega) = \beta \omega \mathcal{S}_{\text{quantum}}(\mathbf{q},\omega)$.
As such, we appropriately re-scale our our classical MD results by a factor of $\beta\omega$ in the subsequent sections.
\section{Relation between the XYZ model and original DO model}

The XYZ model introduced in the main text is obtained from the general symmetry-allowed pseudospin-1/2 model,
\begin{equation}
H=\sum_{\langle i,j\rangle}\left[J_{xx}\tau_i^x\tau_j^x+J_{yy}\tau_i^y\tau_j^y+J_{zz}\tau_i^z\tau_j^z+J_{xz}(\tau_i^x\tau_j^z+\tau_i^z\tau_j^x)\right]-\sum_i\left[({\bm h}\cdot\hat{z}_i)(\tilde{g}_z\tau_i^z+\tilde{g}_x\tau_i^x)\right],
\end{equation}
where $\hat{z}_i$ denotes the local $z$ direction of the sublattice at site $i$. We introduced here the pseudospin-1/2 operator ${\bm \tau}$: $\tau^x=J_x^3-\overline{J_xJ_yJ_y}$, $\tau^y=J_y^3-\overline{J_yJ_xJ_x}$, and $\tau^z=J_z$, where the overline indicates a symmetric product. Note that $\tau^x$ and $\tau^z$ transform as dipolar moments, and thus can linearly couple to the external magnetic field. The mixing term, i.e., $J_{xz}$ term can be eliminated by performing a pseudospin rotation about the local $\hat{y}_i$ axis by angle $\theta$:
\begin{equation}
    S^x=\tau^x\cos\theta-\tau^z\sin\theta,\quad S^y=\tau^y,\quad S^z=\tau^x\sin\theta+\tau^z\cos\theta.
\end{equation}
Here, $\tan2\theta=2J_{xz}/(J_{zz}-J_{xx})$. This transformation results in the XYZ Hamiltonian, Eq. (2) in the main text with redefined $g$-factors, $g_x=\tilde{g}_x\cos\theta-\tilde{g}_z\sin\theta$ and $g_z=\tilde{g}_x\sin\theta+\tilde{g}_z\cos\theta$. Finally, exchange couplings are also renormalized to be
\begin{equation}
    {\cal{J}}_x=\frac{J_{zz}+J_{xx}}{2}-\frac{\sqrt{(J_{zz}-J_{xx})^2+4J_{xz}^2}}{2},\quad{\cal{J}}_y=J_{yy},\quad{\cal{J}}_z=\frac{J_{zz}+J_{xx}}{2}+\frac{\sqrt{(J_{zz}-J_{xx})^2+4J_{xz}^2}}{2}.
\end{equation}
In this work, we assume that $g_x$ factor to zero, which physically corresponds to negligible mixing of the octupolar and dipolar moment.

\section{Side-by-Side Phase diagrams of dipolar and octupolar dominant regimes}

\begin{figure}[t!]
\includegraphics[width=0.7\columnwidth]{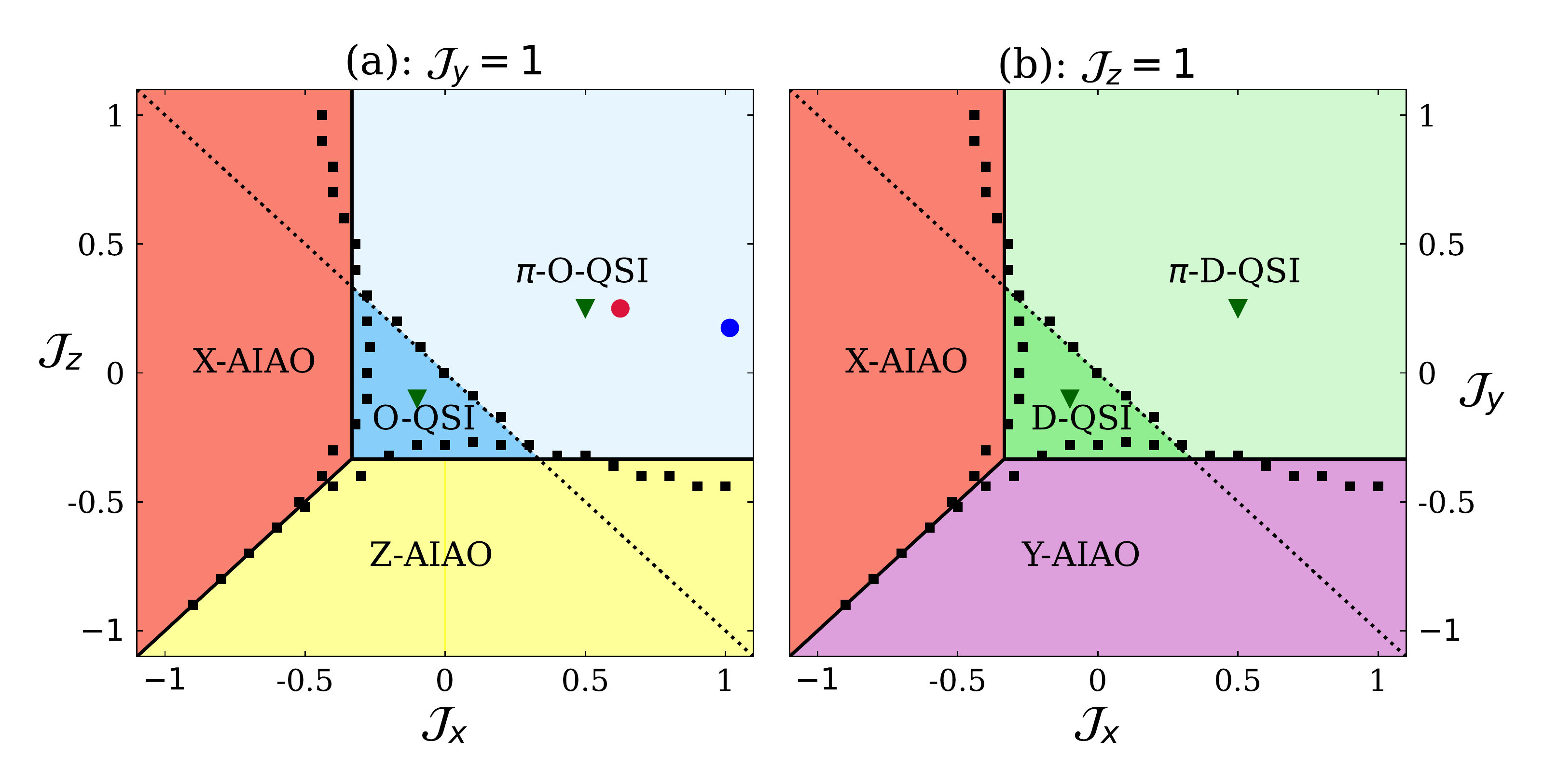}
\caption{\label{pdfig_sm} Side-by-side phase diagram of DO systems in zero magnetic field for the octupolar dominant and dipolar dominant regimes 
As in Fig. 1 of the main text, the X,Y,Z all-in, all-out phases are labelled as X,Y,Z-AIAO, while the $0-$flux ($\pi-$flux) quantum spin ice phases for the corresponding octupolar and dipolar dominant regimes are labelled as O-QSI ($\pi-$O-QSI) and D-QSI ($\pi-$D-QSI). The solid lines indicate the classical phase diagram phase boundaries, while the black squares are the 16-site ED phase boundaries from Ref. \citep{Patri2020}. The blue and red circles are superimposed experimentally extracted parameter sets from Ref. \citep{Smith2021} and \citep{Bhardwaj2021}, respectively.
The inverted-green triangles are the parameter sets employed in the MD and ED studies in the subsequent sections.
The dashed lines indicate the crossover from unfrustrated to frustrated exchange couplings.}   
\end{figure}

We present in Fig. \ref{pdfig_sm} the phase diagram of DO systems in zero magnetic field in the octupolar dominant and dipolar dominant regimes. 
This is the same phase diagram as Fig. 1 in the main text, but with the cube faces placed adjacent to each other for clarity.

\section{Spin polarization analysis}
To understand the signatures of our total neutron scattering results, we include a spin-polarization analysis of the  equal-time neutron scattering structure factors. The structure factor in the NF channel is given by 
\begin{equation}
  S_{\text{NSF}}({\bm q})
  =\frac{1}{N}\sum_{i,j}(\hat{z}_i\cdot{\bm \hat{v}})(\hat{z}_j\cdot{\bm \hat{v}})
  e^{-\mathrm{i}{\bm q}\cdot({\bm R}_i-{\bm R}_j)}\langle S_i^z S_j^z\rangle.
\end{equation}
and by
\begin{equation}
  S_{\text{SF}}({\bm q})
  =\frac{1}{N}\sum_{i,j}(\hat{z}_i\cdot{\bm \hat{v}_\perp})(\hat{z}_j\cdot{\bm \hat{v}_\perp})
  e^{-\mathrm{i}{\bm q}\cdot({\bm R}_i-{\bm R}_j)}\langle S_i^z S_j^z\rangle.
\end{equation}
in the SF channel. Here, the neutrons are polarized in the direction of unit vector $\bm \hat{v}$ perpendicular to the scattering plane, and $\bm \hat{v}_\perp=\frac{\bm \hat{v}\times \bm q}{|\hat{v}\times \bm q|}$. 

\begin{figure*}[h!]
    \includegraphics[width=0.7\columnwidth]{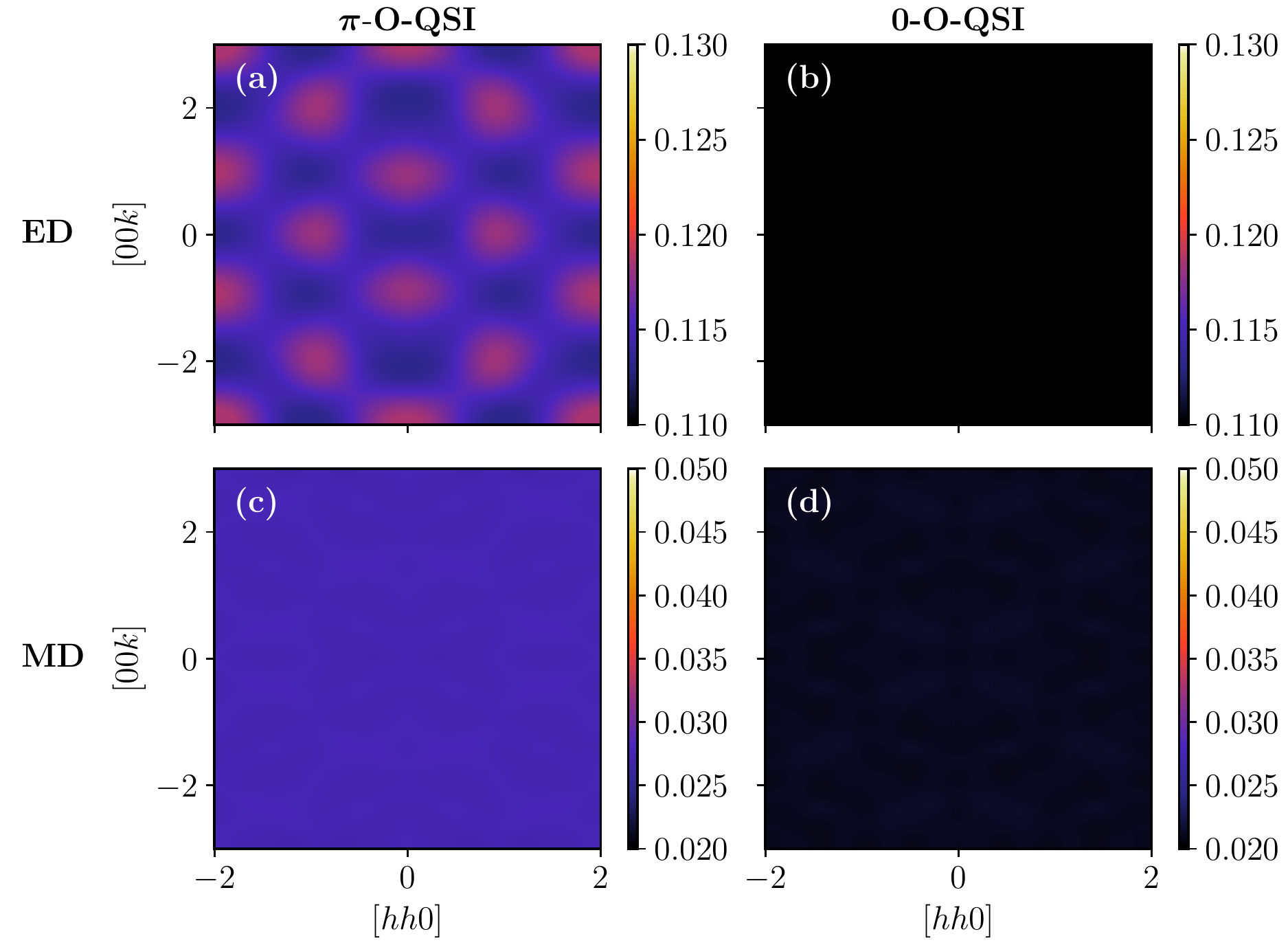}
    \caption{Non-spin flip channel of the equal-time neutron scattering structure factor for the octupolar QSI regimes.}\label{nsf} 
\end{figure*}

\begin{figure*}[h!]
    \includegraphics[width=0.7\columnwidth]{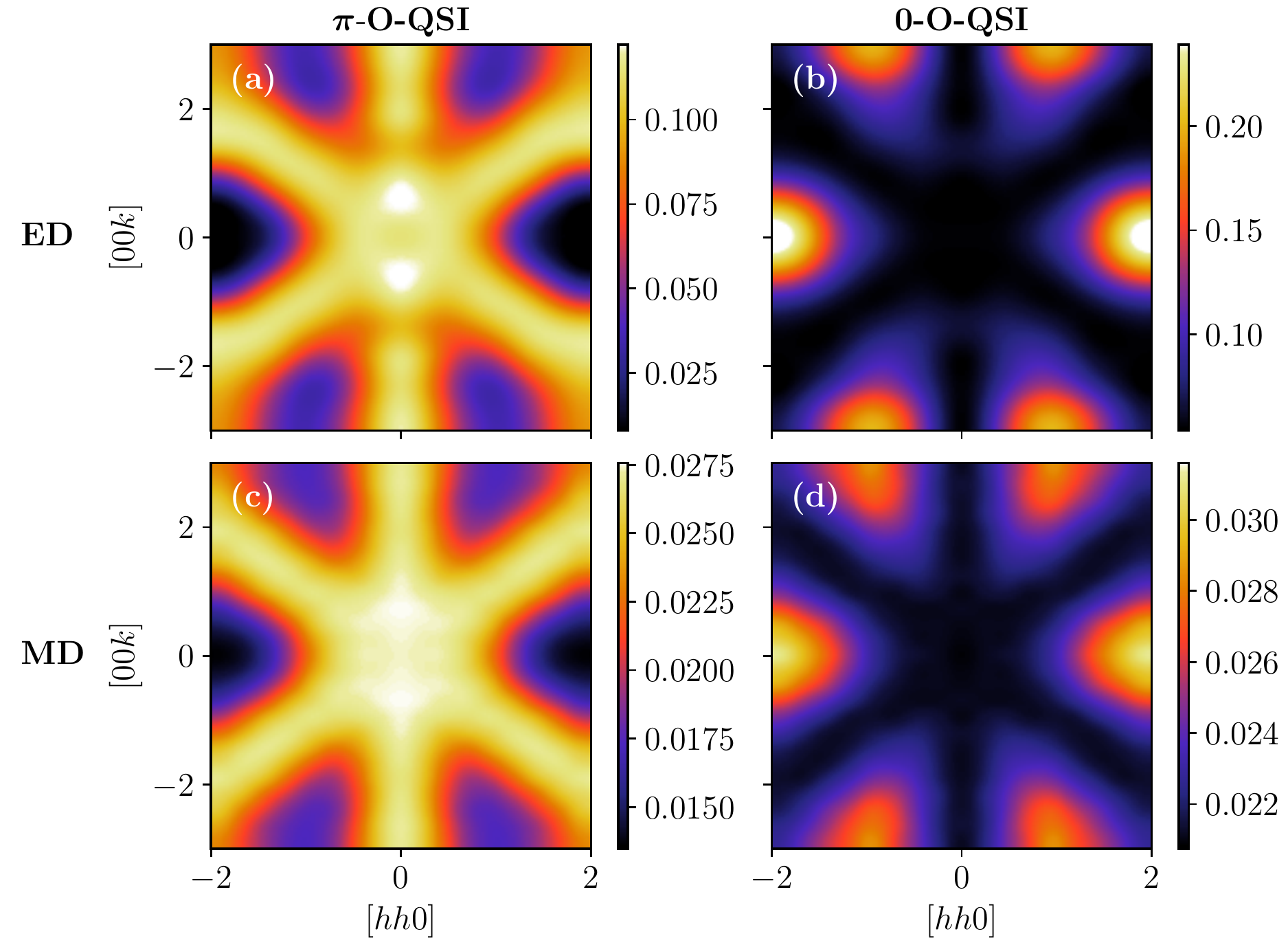}
    \caption{Spin flip channel of the equal-time neutron scattering structure factor for the octupolar QSI regimes.}\label{sf} 
\end{figure*}

Here, we show the equal-time neutron scattering structure factors in the non-spin-flip (NSF) and spin-flip (SF) channels in Figs. \ref{sf} and \ref{nsf} respectively. The most fascinating observation can be made in our NSF results. We observe a weak modulation of the intensity in the quantum calculation of the $\pi$-flux oQSI in Fig. \ref{nsf}(a), whose intensity structure is consistent with those observed in the experiments of Ref.~\citep{Smith2021}. Namely, we observe high intensity peaks at the $[001]$ and $[003]$ (and symmetry equivalent) points, and low intensities everywhere else. We emphasize that this intensity structure is unique to the $\pi$-flux QSI phase. The scattering in the SF channel in this regime in Fig. 3(a), on the other hand, shows the rod-like structures also seen in the total neutron scattering data, with high intensity peaks at the $[001]$ and $[003]$ points, also consistent with Ref.~\citep{Smith2021}. We expect this structure in the NSF channel to arise from fluctuations of the matter fields in the octupolar case, since the photons do not couple to the neutrons here. 

In the classical case, the intensity profile in the NSF channel is largely uniform as expected. Since the classical calculations are performed at finite temperature, the spin correlations are expected to decay exponentially as a function of $T$ at the nearest-neighbour level, whereas they persist in the presence of explicit dipolar interactions\cite{Isakov2005}. The contrast between the classical and quantum pictures in the $\pi$-flux O-QSI regime further convinces us of the role of quantum fluctuations in the neutron scattering experiments. 

\section{Exact diagonalization interpolation scheme}
\begin{figure}
    \includegraphics[width=0.5\columnwidth]{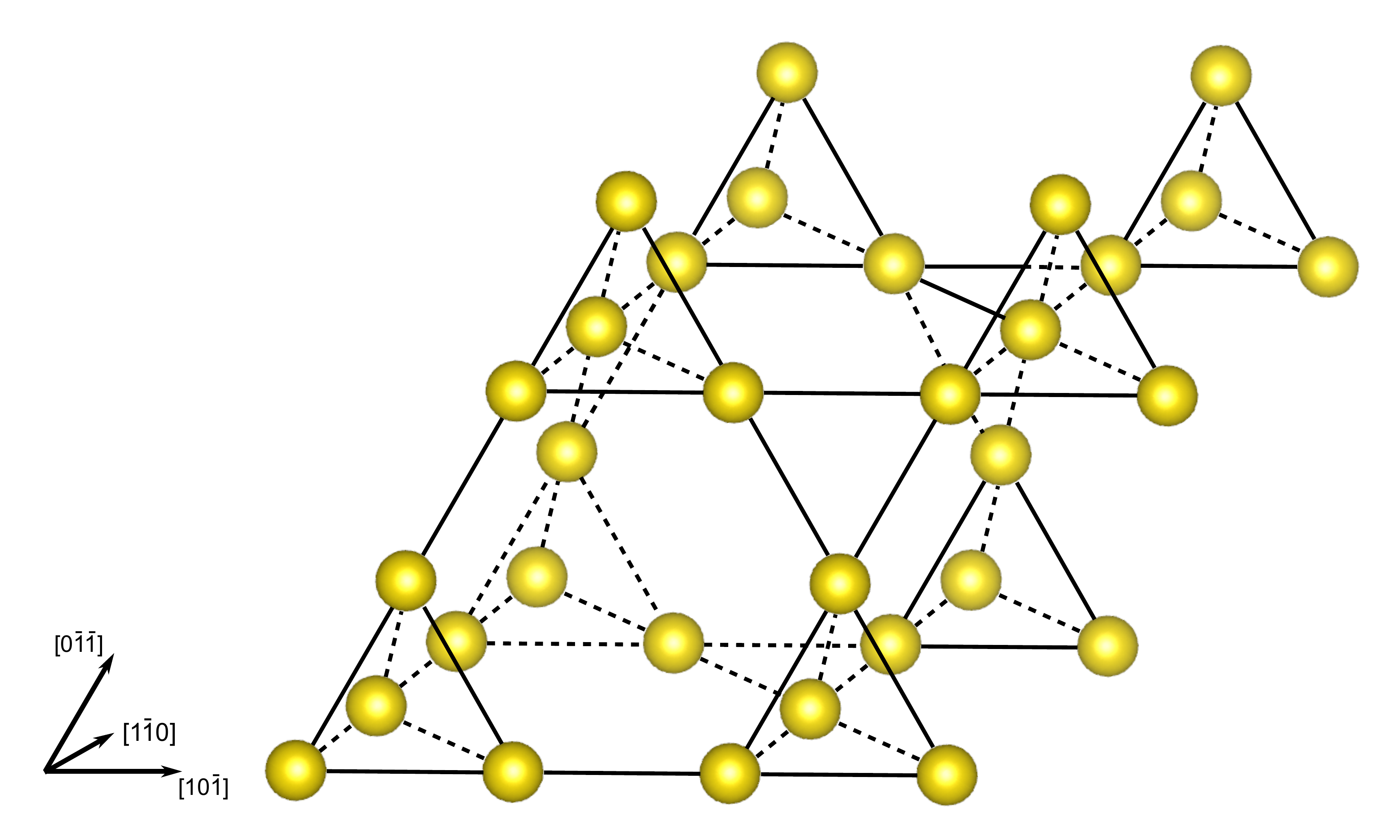}
    \caption{Schematic illustration of the 32-site symmetric cluster of the pyrochlore lattice. The periodic condition is assumed along the crystal $[1\bar{1}0]$, $[10\bar{1}]$, and $[0\bar{1}\bar{1}]$ directions.}
    \label{EDcluster}
\end{figure}

\begin{figure*}[t!]
    \includegraphics[width=\columnwidth]{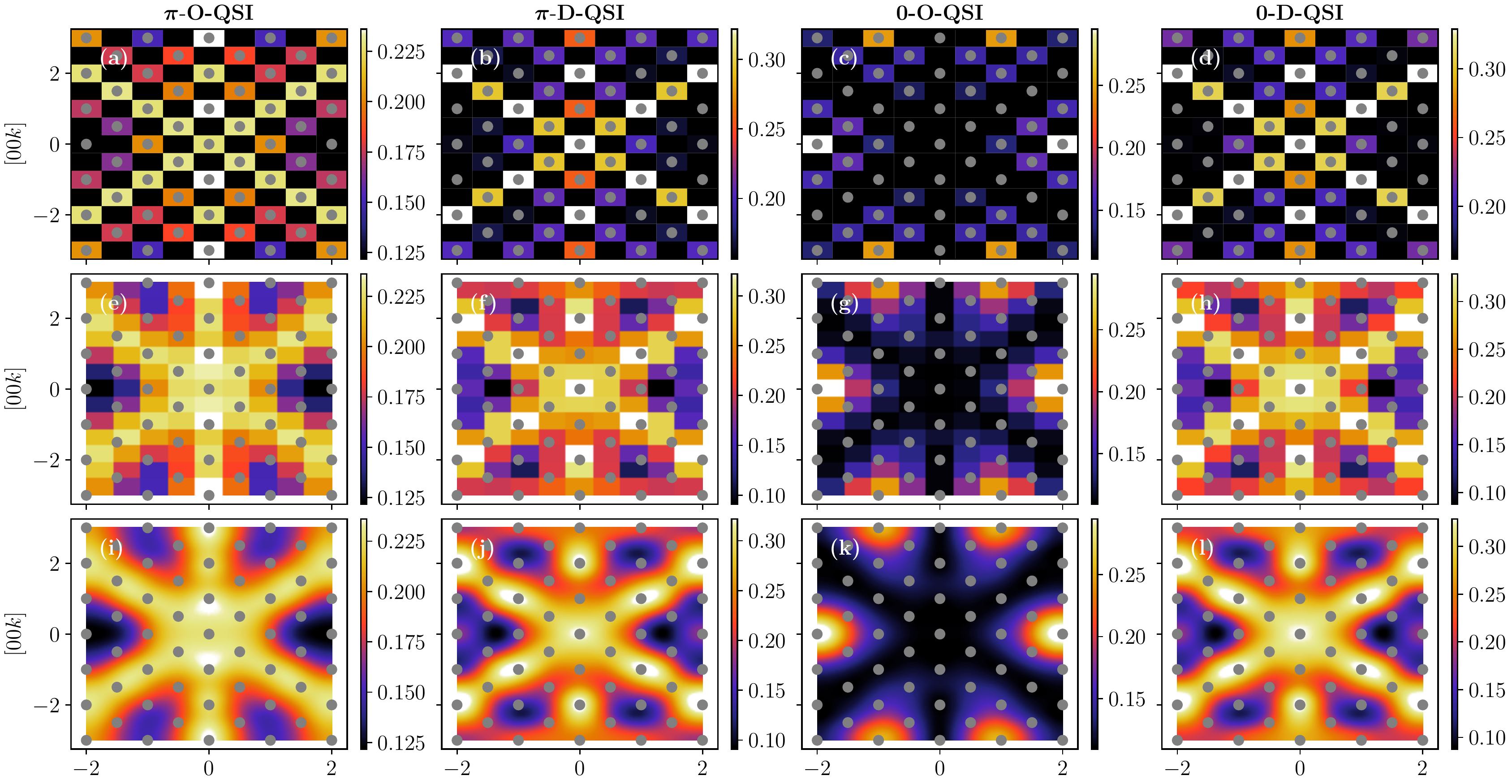}
    \caption{\label{ns-interp} Interpolation scheme for the total neutron scattering results from 32-site exact diagonalization (ED) shown with allowed momentum positions (grey dots). (a)-(d) shows the raw ED data with no interpolation, with missing data points coloured black. (e)-(h) shows the first level of interpolation over the missing data points. (i)-(l) shows the smoothing out of (e)-(h) using a spline36 method.}
    \label{interp}
\end{figure*}
\begin{figure*}[h!]
    \includegraphics[width=\columnwidth]{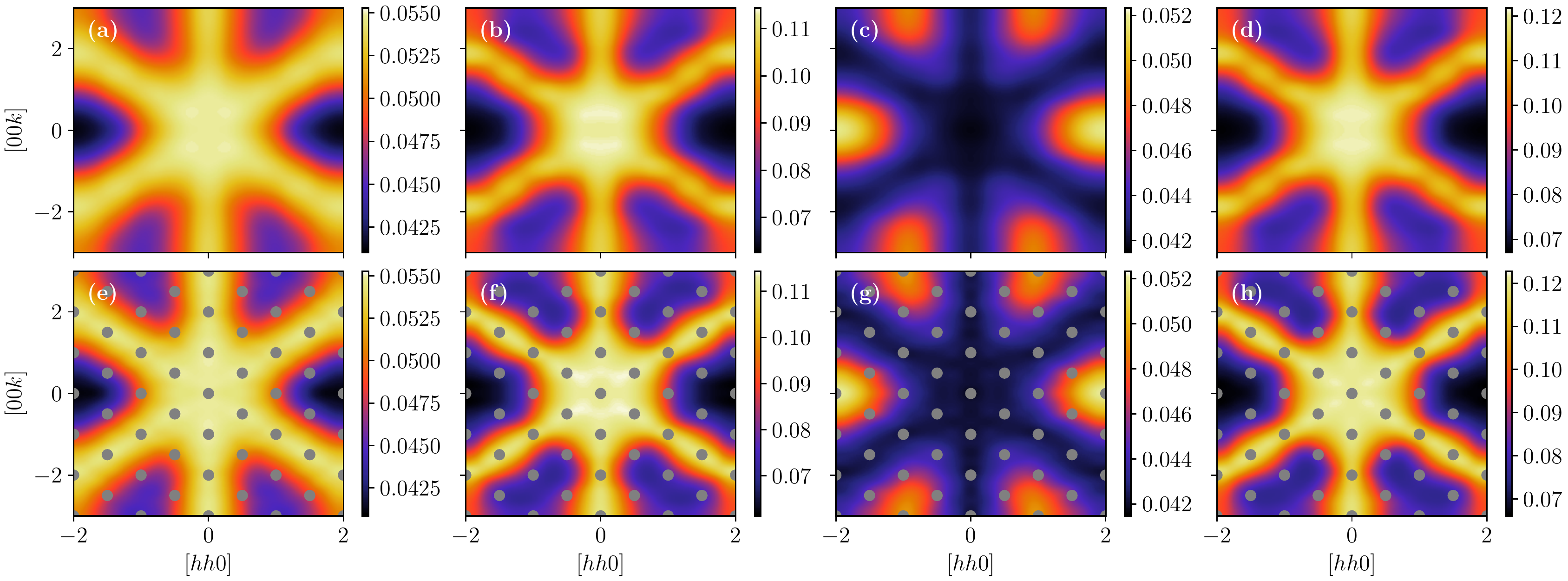}
    \caption{Classical results interpolated over the ED grid using the same interpolation scheme. (a)-(d) shows the data over the original grid, and (e)-(f) shows the data interpolated over the ED grid (shown in grey dots). }\label{interpmd} 
\end{figure*}

In the ED calculation, we employed the 32-site cluster shown in Fig. \ref{EDcluster}, where the periodic condition is assumed along the crystal $[1\bar{1}0]$, $[10\bar{1}]$, and $[0\bar{1}\bar{1}]$ directions \cite{Momma2011}.
Due to the limited number of momentum points accessible using 32-site exact diagonalization, we use an interpolation scheme in order to better compare against the classical results. The accessible points in the $[hhk]$ plane are given by $\mathbf{q}=n\mathbf{a}+m\mathbf{b}$, where $n,m$ are integers, $\mathbf{a}=(\pi,\pi,\pi)$, and $\mathbf{b}=(-\pi,-\pi,\pi)$. Fig. \ref{interp} (a)-(d) shows the data with no interpolation, with the missing data between the momentum points coloured black. From these raw data, we see the main features from which we draw our main observations are present. Fig. \ref{interp} (e)-(h) shows the first level of interpolation, which simply fills in the missing data via a cubic interpolation scheme of a triangular grid, and for artistic purposes, (i)-(l) smooths over this grid via the spline36 interpolation method from Python's matplotlib package.  
As a further sanity check, we apply the same interpolation methods on our classical results, shown in Fig. \ref{interpmd}. We can see that the classical results interpolated over the ED momentum grid yields almost identical plots when compared with the original grid.

\bibliography{U1-Pi}